# Synthesis and magnetic properties of NiFe$_{2-x}$Al$_x$O$_4$ nanoparticles

A.T.Raghavender$^{a,*}$, Damir Pajic$^b$, Kreso Zadro$^b$, Tomislav Milekovic$^b$

P.Venkateshwar Rao$^c$, K.M.Jadhav$^d$, D.Ravinder$^e$

$^a$ Department of Physics, Swami Ramanand Teerth Maratwada University, Nanded, Maharastra-431606. India

$^b$Department of Physics, Faculty of Science, University of Zagreb, Bijenicka c. 32, HR-10000 Zagreb, Croatia

$^c$Department of Chemistry, P.G college of Science, Saifabad, Osmania University, Hyderabad –500004. India

$^d$Department of Physics, Dr.Babasaheb Ambedkar Maratwada University, Aurangabad-431004. India

$^e$Department of Physics, P.G college of Science, Saifabad, Osmania University, Hyderabad –500004. India

**Abstract**

Nanocrystalline Al doped nickel ferrite powders have been synthesized by sol-gel auto-ignition method and the effect of non-magnetic aluminum content on the structural and magnetic properties has been studied. The X-ray diffraction (XRD) revealed that the powders obtained are single phase with inverse spinel structure. The calculated grain sizes from XRD data have been verified using Transmission electron microscopy (TEM). TEM photographs show that the powders consist of nanometer sized grains. It was observed that the characteristic grain size decreases from 29nm to 6 nm as the non-magnetic Al content increases, which was attributed to the influence of non-magnetic Al-concentration on the grain size. Magnetic hysteresis loops were measured at room temperature with maximum applied magnetic field of $\approx 1T$. As aluminum content increases, the measured magnetic hysteresis curves became more and more narrow and the saturation magnetization and remanent magnetization both decreased. The reduction of magnetization compared to bulk is a consequence of spin non-collinearity. Further reduction of magnetization with the increase of aluminum content is caused by non-magnetic $Al^{3+}$ ions and weakened interaction between sublattices. This, as well as the decrease in hysteresis was understood in terms of the decrease in particle size.

**Key points:**

Sol-gel, nanoparticles, grain size, magnetic properties, Ni-ferrite, spin non-collinearity.

**Corresponding Author:**

Tel:- +919440532011

e-mail: **raghavi9@gmail.com** (A.T.Raghavender)

1. **Introduction**

Synthesis of nanometer size particles proved to be one of the interesting fields of material science in material processing and technological applications, as the small size particles have some of the interesting properties as compared to bulk particles. These particles have improved catalytic, dielectric and magnetic, properties, as they possess high resistivity and negligible eddy current losses [1-5]. The preparation technique also plays an important role in surface properties and the Curie temperature ($T_c$) can also be varied by substitution of non-magnetic cations. Magnetic nanoparticles promise some interesting applications, such as in high frequency devices, magnetic fluids, high density magnetic recording, colour imaging etc [6-8].

The various processing techniques, which are used for the synthesis of spinel ferrite powders include microwave refluxing [9], sol-gel [10-13], hydrothermal [14, 15], co-precipitation [16], spray pyrolysis [17]. In fact there are numerous papers on synthesis of nickel ferrite by various methods. In the present investigation we have employed sol-gel auto-ignition method to synthesize Al doped nickel ferrite nanoparticles. The sol-gel auto-ignition method is used to speed up the synthesis of complex materials. It is a simple process, which offers a significant saving in time and energy consumption over the traditional methods, and requires less sintering temperature. This method is employed to obtain improved powder characteristics, more homogeneity and narrow particle size distribution, thereby influencing structural, electrical, and magnetic properties of spinel ferrites.

In the inverse spinel structure of $NiFe_2O_4$ the tetrahedral sites are occupied by ferric ions and octahedral by ferric and nickel ions. The investigation of aluminum substituted nickel ferrite $NiFe_{2-x}Al_xO_4$ was not well documented and we present here the synthesis, structural characterization and also the basic magnetic properties of $NiFe_{2-x}Al_xO_4$ nanoparticles.

## 2. Experimental Procedure

### 2.1 Synthesis technique

Nanocrystalline powders of $NiFe_{2-x}Al_xO_4$ (x = 0.0,0.2,0.4,0.6,0.8,1.0) were prepared by sol-gel auto-ignition method. The A.R Grade Citric acid ($C_6H_8O_7.H_2O$), Nickel Nitrate ($Ni(NO_3)_2.6H_2O$), Ferric Nitrate ($Fe(NO_3)_3.9H_2O$), Aluminum Nitrate($Al(NO_3)_3.9H_2O$) (>= 99%) were used as starting materials. The molar ratio of metal nitrates to citric acid was taken as 1:3. The metal nitrates were dissolved together in a minimum amount of de-ionized water to get a clear solution. An aqueous solution of citric acid was mixed with metal nitrates solution, then ammonia solution was slowly added to adjust the pH at 7. The mixed solution was moved on to a hot plate with continuous stirring at $90^0$C. During evaporation, the solution became viscous and finally formed a very viscous brown gel. When finally all remaining water was released from the mixture, the sticky mass began to bubble. After several minutes the gel automatically ignited and burnt with glowing flints. The decomposition reaction would not stop before the whole citrate complex was consumed. The auto ignition was completed within a minute, yielding the brown-colored ashes termed as a precursor. The as-prepared powders of all the samples were heat treated separately at $500^0$C for four hours to get the final product.

**2.2 X-ray diffraction studies**

The structural characterization of the ferrite powders as prepared was carried out using Inel X-ray diffraction system with Ni filter using Co –Kα radiation (wave length λ = 1.78894 $A^o$). The average particle size D was calculated using most intense peak (311) employing the Scherrer formula [18]

$$D = \frac{0.9\lambda}{\beta \cos\theta} \qquad (1)$$

where β is the angular line width at half maximum intensity and θ the Bragg angle for the actual peak.

**2.3 TEM studies**

To verify the particle sizes calculated using XRD data, TEM studies were carried out using Philips CM-12 Transmission Electron Microscope (TEM).

**2.4 Thermal studies**

In order to investigate the spinel ferrite phase formation, the dried powders were characterized via thermogravimetric (TG) and differential thermal analysis (DTA) using Netzsch STA 409 TG-DTA instrument, at a heating rate of $10^0$ C/min in static air.

**2.5 Magnetic measurements**

Magnetic measurements were performed using the commercial PARC EG&G vibrating sample magnetometer VSM 4500. Magnetic hysteresis loops were measured at room temperature with maximal applied magnetic fields up to 0.95T. Magnetic field sweep rate was 5 Oe/s for all measurements, so that the measurement of hysteresis loops with maximum field of 0.95T took about three hours. The samples prepared in powder form were fixed in paraffin in order to exclude the motion of powder in a measuring cap. The saturation magnetization, coercivity and remanent magnetization were found from hysteresis loops.

**3. Results and Discussion**

**3.1 Structural characterization**

Figure 1 shows the X-ray diffractograms of $NiFe_{2-x}Al_xO_4$ ( x = 0.0, 0.2, 0.4, 0.6, 0.8, 1 )samples. The XRD patterns clearly indicate that the prepared samples contain cubic spinel structure only [19]. A close examination of XRD patterns reveal that the diffraction peaks became broader with increasing aluminum content *x*, which may be due to distribution of nanocrystallinity. The sizes of crystallites in the sample were evaluated by measuring the FWHM of the most intense peak (311). The results are as shown in table 1.

Further, it is observed from table 1 that particle size decreases with increase in non-magnetic Al substitution. The same trend in variation of crystallite size was observed in Mn-Zn ferrite ferrofluids prepared by chemical synthesis [20].

The values for lattice constants were obtained for all the samples using XRD data with an accuracy of ± 0.002 A$^O$. The values of lattice constants are listed in table 1. It is observed from table 1, that the lattice constant decreases with increasing aluminum content *x*. This behaviour of lattice constant with aluminum content *x* is explained on the basis of difference in ionic radii of $Fe^{3+}$ and $Al^{3+}$. In the present series $NiFe_{2-x}Al_xO_4$, larger $Fe^{3+}$ (0.67A$^o$) ions are replaced by smaller $Al^{3+}$ (0.51A$^o$) ions; therefore decrease in lattice constant takes places. Similar trend was found in Cu-Cd ferrite with aluminum substitution [21].

The addition of $Al^{3+}$ ion which has strong preference for the octahedral sites should exhibit the decrease of the magnetization because the ion is non-magnetic. It was observed [21] that increasing the amount of Al, the magnetic moment of unit cell decreases. They have found that the canting angle increases with increase of the Al content, which indicates the favouring of triangular spin moment at the octahedral site leading to reduction in the sublattice interaction. The removal of magnetic $Fe^{3+}$ ion from magnetic sublattice and substitution of the non-magnetic $Al^{3+}$ ion in its place weakens the superexchange interactions, which tend to align the neighbouring dipoles antiparallelly. Hence, the structural and magnetic properties are closely interconnected.

Figures 2 (a) and (b) present the TEM photographs of x = 0.0 and 1.0, which show that the particle sizes in nanometers agree well with XRD data.

**3.2 Thermal Analysis**

In order to investigate the formation of the spinel structure phase, thermal analysis for the sample x= 0.0 is carried out in the temperature range $100^0$C to $500^0$C in static air at $10^0$ C/min. Fig 3 shows TG – DTA for precursor cobalt ferrite powders. The TG-DTA shows the presence of one exothermic peak at $303^0$C which may be due to reaction of citric acid and metal nitrates with total weight loss around 6%. The exothermic peak at $303^0$C indicates the formation of crystallization of the ferrite phase. It may be mentioned that the formation of the spinel phase is at a lower temperature than $303^0$C while thermal analysis is a dynamic process. Thus the temperature recorded by TG-DTA for spinel formation is expected to be higher. It is observed from acetate-citrate gelation method [22], that above $430^0$C only the organics are removed completely, and the prediction of particle size and further analysis is irrelevant below this temperature. Therefore even though the decomposition has completed at $303^0$C, we have chosen $500^0$C as final calcination temperature for this samples. The preparation of ferrites around this temperature is confirmed by XRD data.

## 3.3 Magnetic properties

Magnetic hysteresis loops of aluminum doped nickel ferrite $NiFe_{2-x}Al_xO_4$ (x = 0.0, 0.2, 0.4, 0.6, 0.8, 1) nanoparticles measured at room temperature using VSM are shown in figure 4. As observed from figure 4 the hysteresis loops become more and more narrow as the aluminum content $x$ increases. At maximal applied magnetic field of 0.95T the saturation was not achieved. However, it is obvious that the magnetization at this field is decreasing monotonously as $x$ increases.

The value of magnetization at applied magnetic field of 1T for our $NiFe_2O_4$ nanoparticles was measured to be $28 Am^2kg^{-1}$. It is considerably lower than the value of $50 Am^2kg^{-1}$ corresponding to bulk sample at room temperature [24]. This lowering of magnetization in fine magnetic particles of $NiFe_2O_4$ was explained previously by non-collinear spin structure induced by several reasons. The results of our measurements presented in table 1 and figure 5 show that with increasing aluminum content $x$ the magnetization $M$ (at the same outer conditions), remanent magnetization $M_r$, saturation magnetization $M_{sat}$ and coercive field $H_c$ all decrease monotonously. $M_{sat}$ is determined by fitting the high field reversible part (above 0.7T) of measured $M(H)$ data by the well known Langevin dependence (equation.2)

$$M(H) = M_{sat} \cdot \left(1 - \frac{k_B T}{M_{sat}\, \rho\, V_{eff}\, \mu_o\, H}\right) \qquad (2)$$

where $T$ is temperature, $\rho$ is density of material, $V_{eff}$ the effective value of particles volume, and the other symbols have usual meaning.

A model of core with the usual spin arrangement and a surface layer with atomic moments inclined to the direction of the net magnetization was confirmed relatively long ago by Mössbauer spectroscopy [24]. Nowadays, there are more evidences of change in magnetization with the change of nanoparticles size, with a general conclusion that in smaller particles the reduction of magnetization is more pronounced [25]. Magnetization reduction was connected to cation redistribution (reduced concentration of iron cations on tetrahedral sites) and to spin non-collinearity [26]. On the other hand, the saturation magnetization can be improved by increasing the crystallinity even in really small particles [27]. Finite size effects and surface phenomena in addition to the changes in the inversion degree inspired some authors to further confirm the mentioned explanation of the change in magnetic properties [28]. Therein, mixed spinel structure in $NiFe_2O_4$ was found using in-field Mössbauer, magnetization, and EXAFS measurements when the grain size is reduced to a few nanometres. Besides the canting of the surface spins caused by broken exchange bonds, the core spins could also have canted spin structure due to the large magnetocrystalline anisotropy resulting from the occupation of the tetrahedral sites by $Ni^{2+}$ ions [29]. Also, smaller magnetization in nanoparticles than in bulk material was explained by spin disorder on surface layer [30]. Another study showed that the spin-canting is responsible for the decrease of magnetization and that the canting is not a surface effect, but it is finite size effect [31]. However, the value of magnetization in our sample is in agreement with previously published magnetization measurements of $NiFe_2O_4$ nanoparticles [24, 25, 29, 30] and with the magnetization of non-annealed milled samples [26]. Therefore, it is possible to understand the observed magnetization reduction in our $NiFe_2O_4$ nanoparticles compared to the bulk magnetization under the light of cited works.

Coercive field of $\mu_o H_c = 23$ mT for our $NiFe_2O_4$ nanoparticles at room temperature is in accordance with the results in [25]. For bigger particles the larger coercivity was reported [24]. This is expected from behaviour of superparamagnetic particles magnetization blocked by magnetic anisotropy [32] and also from the Stoner-Wohlfarth model [33]. In our present investigation the coercivity is much lower than in the milled nanoparticles [26] of approximately same sizes. There the coercivity decreases with annealing and our value is comparable to coercivity of sample annealed at highest temperature. This shows that in our samples higher order is produced during the synthesis, while this order is accessible after the strong annealing of surface disordered particles produced by milling.

Systematical doping of some spinel ferrite with aluminum or any other element and the underlying study of magnetic properties is not extensively documented. Substitution of iron with aluminum was performed in Ni-Zn ferrite [34] in the form of bulk. There the magnetic moment and ratio of remanence and saturation magnetization both decrease. It is explained by the exchange interaction and the effect of magnetic anisotropy [34]. The decrease of magnetic moment with increasing Al content suggests that canted spins occur in the system. The exchange interaction between the sublattices reduces as $Fe^{3+}$ content decreases, reducing the magnetization, too. Therefore, many reasons lead to the reduction of magnetization measured in our samples.

In our investigation it was observed that addition of Al ions results in decrease of the grain size. The same was observed by other researchers [21, 34]. The coercive field decreases generally with decreasing of particle size, because of lower anisotropy barriers, which are in first approximation proportional to the volume of nanoparticles [32, 33]. The dependence of coercive field on particle size *d* is shown in figure.6 that should be interpreted carefully because the amount of Al is not constant. In our case, the coercivity decrease is the combination of the size reduction and the mentioned effect of reduced interaction after the increase of Al content. The same arguments are valid for the remanent magnetization dependent on *x* and *d* shown in figure 6, too.

The doping of cobalt ferrite nanoparticles with lanthanide elements [35] shows that the consequences could be rather complex but very interesting matter concerning the blocking temperature, coercivity, saturation magnetization, remanence and other properties. Also, it is shown that using the bare strength of spin-orbital coupling cannot satisfactorily explain the doping effect on magnetic properties [35]. In barium ferrite, after the replacement of iron ions by small amounts of Zn and Sn progressively induced slight decrease of saturation magnetization and relatively larger decrease of coercive field and anisotropy density [36]. The complexity of magnetic behaviour in our samples demands more detailed investigation postponed for the future.

## 4. Conclusions

The sol-gel auto-ignition technique yields nanocrystalline single phase ferrites. The XRD pattern shows the formation of single phase cubic spinel structure for all the samples. The lattice constant decreases with Al concentration. Thermal analysis studies indicate the formation of spinel phase at $350^0$C. TEM analysis confirmed that the particle size decreases as the non-magnetic Al content increases. Measured magnetic properties including the room temperature magnetic hysteresis loops of our samples show two kinds of results. First, the reduction of magnetization because of nanoparticulated structure in comparison to bulk material is in accordance with previously documented investigations. The reduction comes from the non-collinearity induced by finite size and surface effects. Second, the decrease of coercive field, saturation magnetization and remanent magnetization with increasing aluminum content occurs because the replacement of $Fe^{3+}$ by $Al^{3+}$ ions weaken the sublattice interaction and lowers the magnetic moments of unit cells. Also, this decrease is related to the decrease of particle size as the aluminum content is increased. Further investigation should give more quantitative description of measured behaviour connected with microstructure.

**Acknowledgements:**

One of the authors A.T.Raghavender, would like to thank Sri.N.Sivaramakrishna, Technical officer, SAIF, IIT Madras for the experimental facilities, Prof.C.Bansal, Prof.Ashok Chatterjee and Sri.Ravi shankar of School of Physics, Central University, Hyderabad for XRD measurements and Prof.D.R.S.Somayajulu, Chairman, Electrical and Electronic Communication Engineering, Institute of Aeronautical Engineering, Hyderabad, India, for valuable discussions.

**Reference:**

[1] H.Gleiter, Prog.Mat.Sci. **33** (1989) 223.

[2] J.Smit and H.P.J.Wijn, "Ferrites" Physical Properties of Ferrimagnetic Oxides in relation to their technical applications. Eindhoven: Phillips (1959).

[3] V.A.M.Barbes, in: Progress in Spinel Ferrite Research, H.K.J.Buschow (Ed), Elseveir, Amsterdam, 1995.

[4] R.J.Rennard, W.L.Khel, J.Catal. **21** (1971) 282.

[5] I.E.Candlish, B,H.Kim, nanostruct.Mater. **1** (1992) 119.

6] M.P.Sharrock, IEEE Trans Magn mag. **2**(1996) 707.

[7] J.Popple Well, L.Sakhnini, J.Magn.Magn.Mater. **149** (1995) 72.

[8] M.Kishimoto, Y.Sakurai, T.Ajima, J.Appl.Phys. **76** (1994) 7506.

[9] Jyotsendu Giri, T.Sriharsha, D.Bhadur, J.Mater.Chem. **14** (2004) 875.

[10] Mathew George, Asha Mary John, Swapna S.Nair, P.A.Joy, M.R.Anantharaman, J.Magn.Magn.Mater. **302** (2006) 190.

[11] H.Sato, T.Umeda. Trans. **34** (1993) 76.

[12] R.H.Arendt, J.Solid State Chem. **8** (1973) 339.

[13] S.Giri, S.Samanta, S.Maji, S.Gangli, A.Bhaumik, J.Magn.Magn.Mater. **288** (2005) 296.

[14] Simon Thompson, Neil J.shirtcliffe, Eoins. O'Keefe, Steve Appleton, Carole C.Perry, J.Magn.Magn Mater. **292** (2005) 100.

[15] Jun Wang, Mater.Sci. Eng. B. **127** (2006) 81.

[16] A.S.Albuquerque, J.D.Ardisson, W.A.A.Macedo, J.L.Lopez, R.Paniago, A.I.C.Persiano, J.Magn.Magn.Mater. **226** (2001)1379.

[17] S.Z.Zhang, G.L.Messing, J.American.Ceramic.Soc. **73** (1990) 61.

[18] H.P.Klug, L.E.Alexander, X-Ray Diffraction procedures for Polycrystalline and Amorphous materials, Wiley, New York, NY, 1997, **P.637**.

[19] ASTM 10-325 (Ni - ferrite) Nat.Bur.Stands (U.S) cir. 539 – 1044.

[20] E.Auzans, D.Zins, E.Blums, R.Massart, J.Mater.Sci. **34** (1999) 1253.

[21] S.S.Suryawanshi, V.V.Deshpande, U.B.Deshmukh, S.M.Kabur, N.D.Chaudhari, S.R.Sawant, Mat. Chem. Phys. **59** (1999) 199.

[22] D.Mondelaers, G.Vanhoyland, H.Van den Rul, J.D'Haen, M.K.Van Bael, J.Mullens, L.C.Van Poucke, Mater.Res.Bulletin. **37** (2002) 901.

[23] J.Chappert, R.B.Frankel, Phys. Rev. Lett. **19** (1967) 570.

[24] A.H.Morrish, K.Haneda, J. Appl. Phys. **52** (1981) 2497.

[25] M.George, A.M.John, S.S.Nair, P.A.Joy, M.R.Anantharaman, J. Magn. Magn. Mater. **302** (2006) 190.

[26] V.Sepelak, D.Baabe, D.Mienert, D.Schultze, F.Krumeich, F.J.Litterst, K.D.Becker, J. Magn. Magn. Mater. **257** (2003) 377.

[27] J.Wang, Mater. Sci. Eng. B. **127** (2006) 81.

[28] S.J.Stewart, M.J.Tueros, G.Cernicchiaro, R.B.Scorzelli, Sol. Stat. Com. **129** (2004) 347.

[29] C.N.Chinnasamy, A.Narayanasamy, N.Ponpandian, K.Chattopadhyay, K.Shinoda, B.Jeyadevan, K.Tohji, and K.Nakatsuka, Phys. Rev. B. **63** (2001) 184108.

[30] R.H.Kodama, A.E.Berkowitz, E.J.McNiff, Jr. and S.Foner, Phys. Rev. Lett. **77** (1996) 394.

[31] F.T.Parker, M.W.Foster, D.T.Margulies, A.E.Berkowitz, Phys. Rev. B. **47** (1993) 7885.

[32] C. P. Bean and J. D. Livingston, J. Appl. Phys. **30** (1959) 120S.

[33] E. C. Stoner, E. P. Wohlfarth, Phil. Trans. Roy. Soc. London A240 (1948) 599; reprinted in IEEE Trans. on Magnetics. **27** (1991) 3475.

[34] A.M.Sankpal, S.S.Suryavanshi, S.V.Kakatkar, G.G.Tengshe, R.S.Patil, N.D.Chaudhari, S.R.Sawant, J. Magn. Magn. Mat. **186** (1998) 349.

[35] M.L.Kahn, Z.J.Zhang, Appl. Phys. Lett. **78** (2001) 3651.

[36] C.K.Ong, H.C.Fang, Z.Yang, Y.Li, J. Magn. Magn. Mater. **213** (2000) 413.

Figure

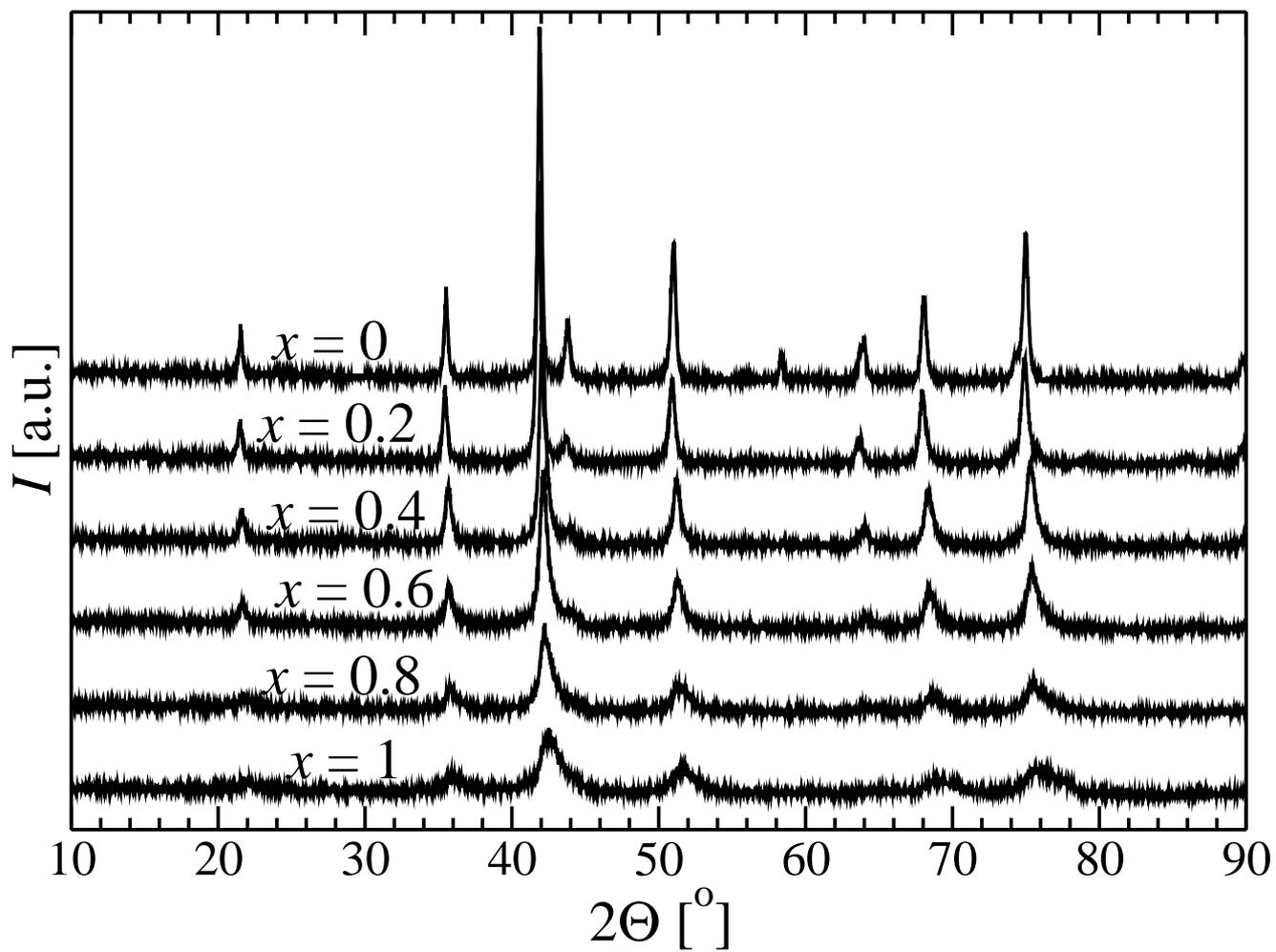

**Figure 1**

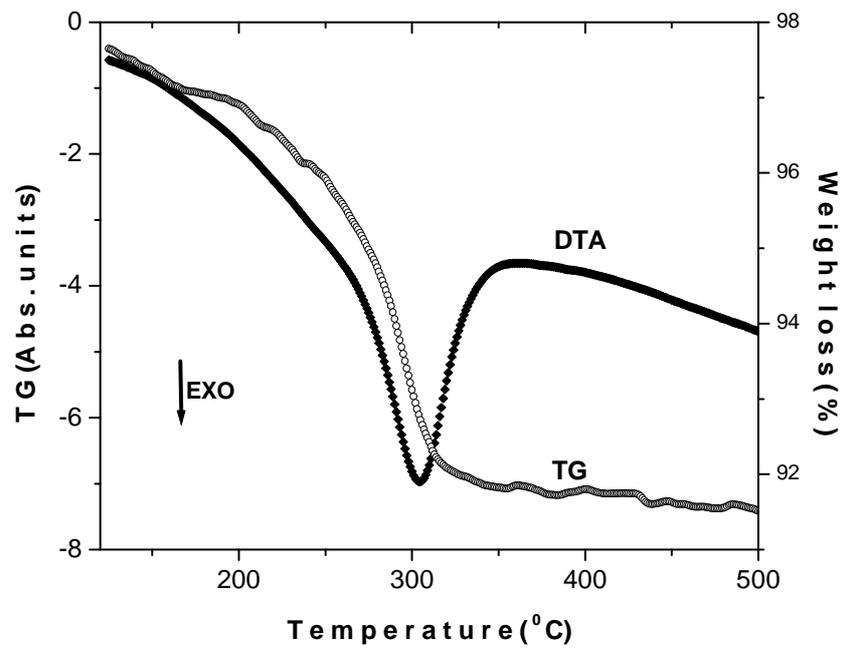

**Figure 3**

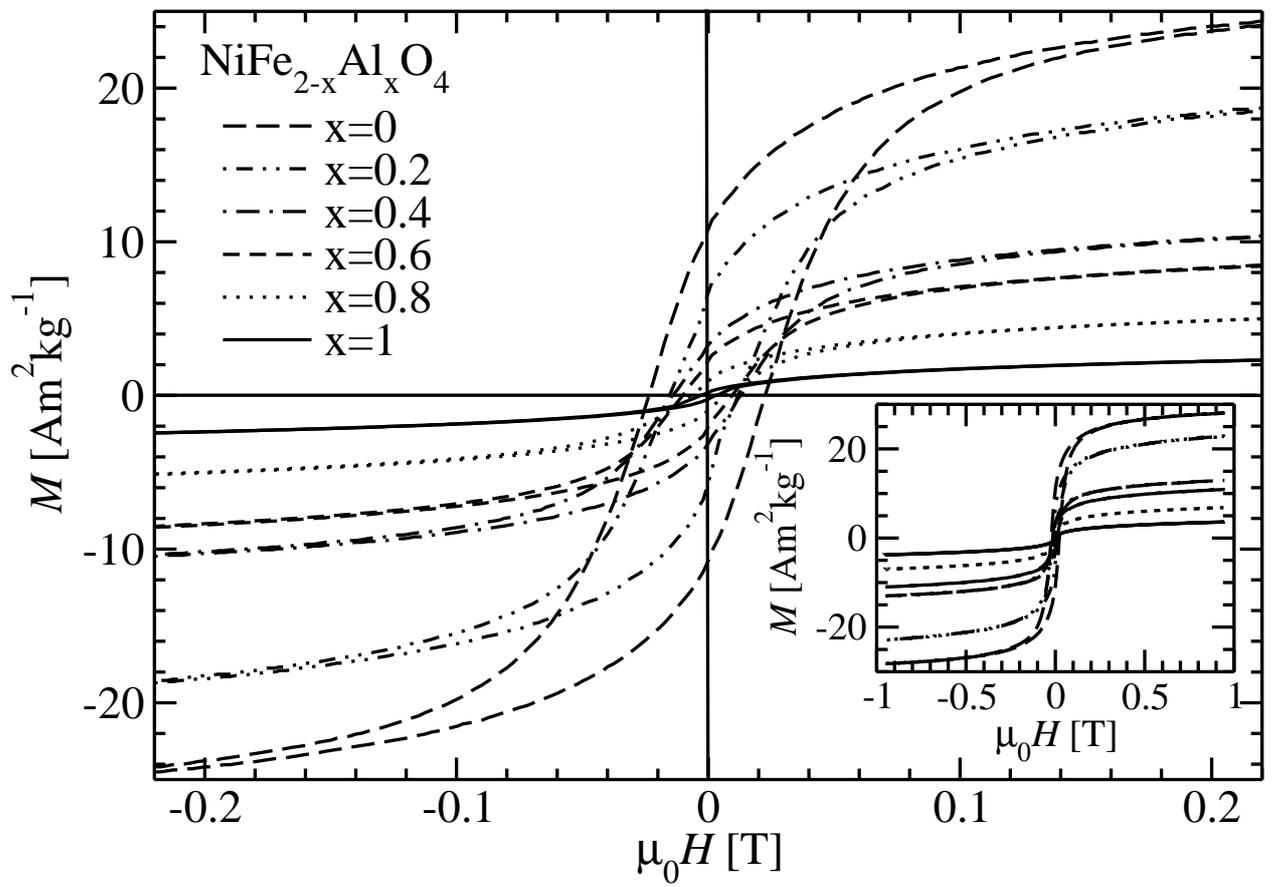

**Figure 4**

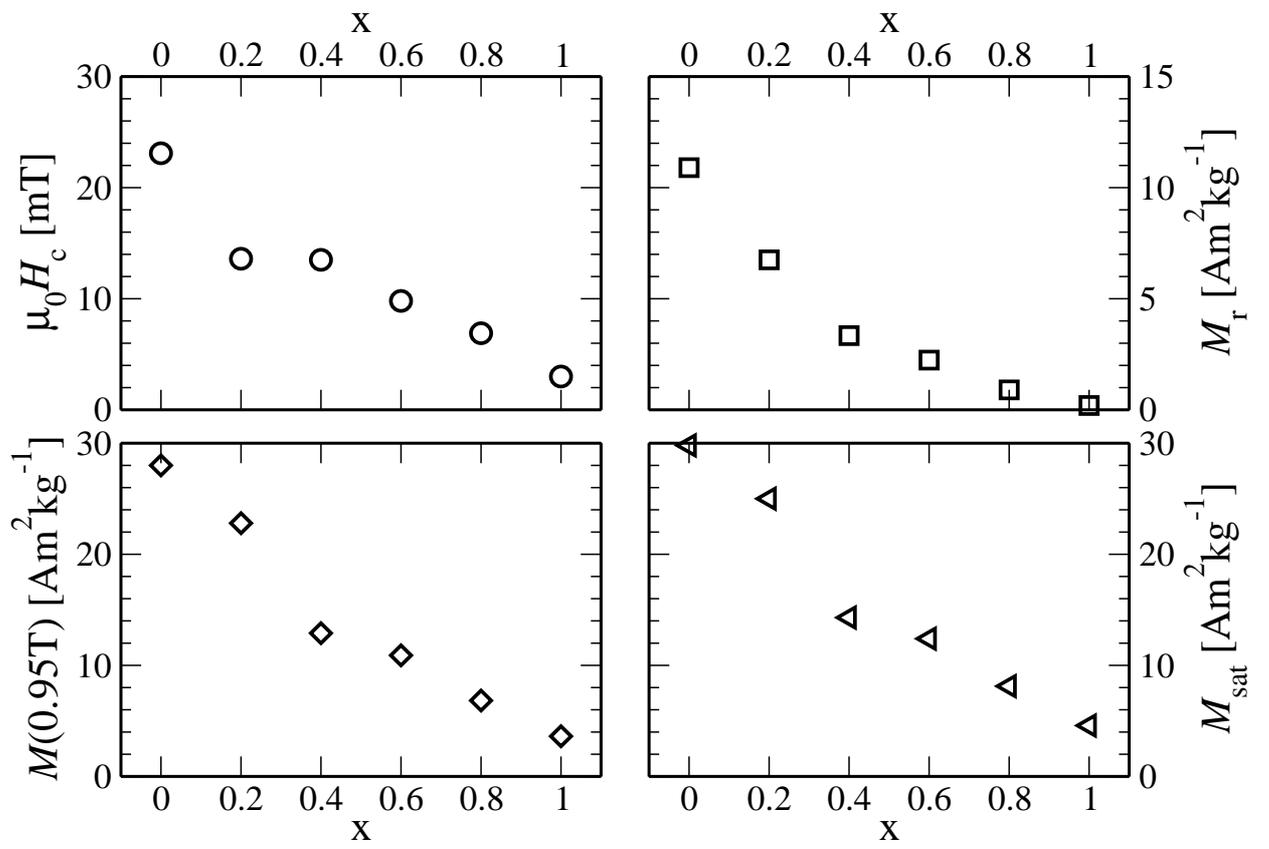

**Figure 5**

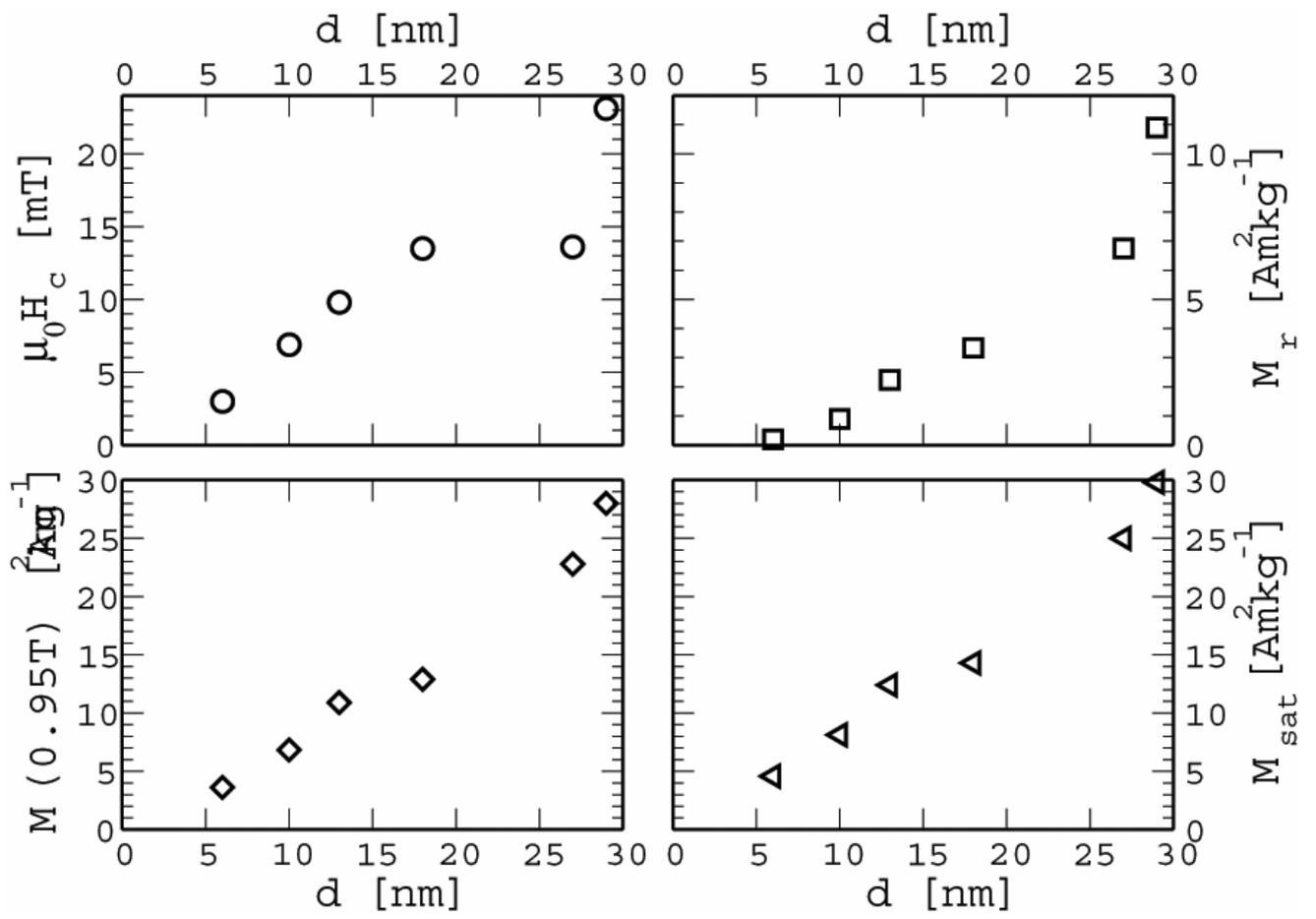

**Figure 6**

**Figures Captions**

**Fig 1.**: X-ray diffraction pattern of NiFe$_{2-x}$Al$_x$O$_4$ (x = 0.0, 0.2, 0.4, 0.6, 0.8, 1.0) nanoparticles.

**Fig 2.**: TEM photographs for samples (a) x = 0.0 and (b) x = 1.0.

**Fig 3.**: TG-DTA of NiFe$_2$O$_4$.

**Fig 4.**: Magnetic hysteresis loops of NiFe$_{2-x}$Al$_x$O$_4$ (x = 0.0, 0.2, 0.4, 0.6, 0.8, 1.0) nanoparticles.

**Fig 5.**: Dependence of coercive field $H_c$, remanent magnetization $M_r$, magnetization at *0.95T M(0.95T)* and saturation magnetization $M_{sat}$ on Al content *x*.

**Fig 6.**: Dependence of coercive field $H_c$, remanent magnetization $M_r$, magnetization at *0.95T M(0.95T)* and saturation magnetization $M_{sat}$ on average particle size *d* of different samples.

**Table.1**

| $x$ | $d$ [nm] | $a$ [A°] | $\mu_o H_c$ [mt] | $Mr$ [Am$^2$/kg] | $M(0.95T)$ [Am$^2$/kg] | $M_{sat}$ [Am$^2$/kg] |
|---|---|---|---|---|---|---|
| 0 | 29 | 8.3024 | 23.1 | 10.9 | 28.0 | 29.8 |
| 0.2 | 27 | 8.2920 | 13.6 | 6.75 | 22.8 | 25 |
| 0.4 | 18 | 8.2685 | 13.5 | 2.34 | 12.9 | 14.3 |
| 0.6 | 13 | 8.2544 | 9.8 | 2.24 | 10.9 | 12.4 |
| 0.8 | 10 | 8.2348 | 6.9 | 0.898 | 6.83 | 8.13 |
| 1.0 | 6 | 8.1821 | 3.0 | 0.194 | 3.62 | 4.58 |

**Table Caption**

**Table 1.:** Dependence of lattice constant $a$, particle size $d$, coercive field $H_c$, remanent magnetization $M_r$, magnetization at 0.95T $M(0.95T)$ and saturation magnetization $M_{sat}$ on Al content $x$.